\documentclass{article}
\usepackage[hyphens]{url}
\usepackage[ocgcolorlinks,pdfusetitle,breaklinks]{hyperref}
\hypersetup{colorlinks=true,breaklinks=true}
\usepackage[bottom]{footmisc}
\usepackage[utf8]{inputenc}
\usepackage{graphicx}
\usepackage{amsmath,mathtools}
\usepackage{slashed}
\usepackage{amssymb}

\usepackage{feynmp-auto}
\usepackage[skip=0.5ex, belowskip=1ex]{subcaption}
\usepackage[export]{adjustbox}
\usepackage[T1]{fontenc}
\usepackage[utf8]{inputenc}
\usepackage{lmodern}
\usepackage[english]{babel}
\usepackage[autostyle]{csquotes}
\usepackage{datetime}
\newdateformat{monthyeardate}{%
  \monthname[\THEMONTH] \THEYEAR}

\usepackage[letterpaper,top=1in,bottom=1in,left=1in,right=1in]{geometry}

\usepackage[numbers]{natbib}   
\usepackage{titlesec}
\titleformat{\section}
  {\normalfont\large\bfseries} 
  {\thesection}             
  {1em}                        
  {}

\title{Existence of Decreasing Nambu Solutions to the Rainbow–Ladder Gap Equation of QCD by Cone Compression} 
\author{Alex Roberts\footnote{
Independent Researcher in San Francisco, Email: alexlewisroberts@gmail.com}}
\date{\monthyeardate\today}

\begin{document}

\maketitle
\abstract{Studying Nambu solutions of the rainbow-ladder gap equation in QCD at zero temperature and chemical potential, we prove that the mass function emerges continuously from zero as the interaction strength is increased past the critical point for all positive, asymptotically perturbative kernels almost everywhere continuous in $L^1$ using the Krasnosel'skii-Guo Cone Compression Theorem. We prove that the coupled system of equations must have a positive, continuous Nambu solution with decreasing mass function for all current quark masses for a class of models which includes the physical point of a popular model of QCD by using a hybrid Krasnosel'skii-Schauder Fixed Point Theorem.}

\section{Introduction}
Establishing the existence of solutions to the gap equation in QCD is not easy; however, the spectrum of solutions has been shown to be rich\cite{Wang_2012} and at finite temperature, the chiral symmetry restoration transition was suggested to be of second-order\cite{H_ll_1999} due to the structure of the fermion propagator which was later verified in numerical simulations\cite{Aoki2006}.\footnote{However, at very small or large quark masses, this is expected to change\cite{guenther2021overviewqcdphasediagram}.} A related phase transition is the emergence of Nambu solutions as the coupling is increased past the dynamical chiral symmetry breaking (DCSB) critical point which we study here at zero temperature and chemical potential.

For a primer on the gap equation for mathematicians and fixed-point theory for physicists, see Appendix \ref{A},\ref{B}. We analyze the solution set of the gap equation after the renormalization scale has been taken to infinity in the equation, see the Appendix for details. Solutions clearly also satisfy a finite renormalization scale equation for some current quark mass parameter but the reverse is not guaranteed. In the case of the gap equation, the current quark mass parameter disappears in the limit of infinite renormalization scale and so any solutions which rely on it to reach an eigenvalue of 1 also disappear with it, namely the Wigner solutions. Solutions that set their own scale as the current quark mass parameter disappears, namely the Nambu solutions, survive and are guaranteed to exist for all current quark masses $m\equiv \lim_{p\rightarrow \infty}(\log \frac{p}{\mu})^{\gamma_m} M(p)$ as we will prove for a popular model using the rainbow-ladder(RL) vertex\cite{Qin_2011}.

Existence results of non-linear integral equations $u(p) = T(u) \equiv \int_0^{\infty} dk K(k,p) f(u(k),k)$, %
often aim to establish a fixed point between two opposite behaviors, say as $\|u\|\rightarrow 0$ and $\|u\|\rightarrow \infty$ with the supremum norm via the Krasnosel'skii-Guo Cone Compression Theorem\cite{Kras,Kwong2008}. The trace of the gap equation takes exactly this form after sending the renormalization scale to $\infty$\cite{Roberts_1994}. %
Another approach called Schauder Fixed Point Theorem can be used if $T(P)\subseteq P$ for some compact input space $P$ as we will see is the case for the trace of $\slashed p$ times the gap equation. Coupled systems with Krasnosel'skii type conditions in both input function have been analyzed in \cite{Infante2015} and a mixed Krasnosel'skii-Schauder theorem has been developed in \cite{INFANTE2024104165}.

Taking the trace of the Euclidean space rainbow ladder gap equation, we define $u(p) \equiv r(p)\frac{M(p)}{Z(p)} \equiv  r(p)B(p)$ for some continuous and positive function $r(p)$ with $\lim_{p\rightarrow\infty}r(p)\sim \log^{\delta}\frac{p}{\mu}$\footnote{We write $\lim_{p\rightarrow p_0} A(p) \sim B(p)$ for $\lim_{p\rightarrow p_0} \frac{A(p)}{B(p)} =  1$.} for some $0 < \delta < \gamma_m$ %
because it allows bounding the gap equation by an integrable kernel and a supremum or infimum over a bounded function.\footnote{For $\gamma_m=0$, any value of $\delta>0$ can be used.} %
In the perturbative region, we show in the appendix that the solution is a sum of the expected $B_{+}(x)=\frac{c_1}{x^{\gamma_m}}$ with $x\equiv \log \frac{p}{\mu}$ and $B_{-}(x) =c_2x^{\gamma_m-1}e^{-2x}$. There, we prove that $c_2$ is always equal to a certain integral and %
we will show that $m\equiv c_1 \geq 0$ is in fact free to choose for a certain guaranteed set of solutions. %
Consider the input space %
$P=\{B(x) \in C[(-\infty,\infty)]\ |\ B(x)\geq0, \exists \gamma_m>0 : \lim_{x\rightarrow \infty}\frac{\partial \log B}{\partial \log x}=-\gamma_m\}$. %
Since $r(p)$ is given, we can switch between $u$ and $B$ whenever convenient and $\lim_{x\rightarrow \infty}\frac{\partial \log B}{\partial \log x}=-\gamma_m$ implies $\lim_{x\rightarrow \infty}\frac{\partial \log T(B)}{\partial \log x}=-\gamma_m$ by \ref{C}.
We have %
\begin{equation}
    f(u,k) = \frac{u}{1+\frac{Z^2(k)u^2}{r^2(k)k^2}}
\end{equation}
and
\begin{equation}
    K(k,p) = \frac{k Z^2(k) r(p)}{4\pi^4 r(k)} \int_{S^3}\mathcal{G}(q^2(k,p))
\end{equation}
with $q_{\mu}=p_{\mu}-k_{\mu}$ with $S^3$ the surface of the sphere that leaves $k^2,p^2$ invariant. We first consider $Z(p)$ to be a known function that one can take to satisfy the gap equation if a solution exists. It should be almost everywhere continuous and converge to a perturbative tail: $Z(p)\rightarrow 1$ as $p\rightarrow \infty$. %
One can write $T=T^Z$ as the results will apply to one unchanging $Z(p)$. We extend our analysis to the coupled system of equations before the conclusion as Schauder's theorem is required to deal with the equation for $Z(p)$. Notice $f(u,k)$ is positive and continuous and we require that $K(k,p)>0$ is continuous almost everywhere, $\int_0^{\infty} K(k,p) dk$ is continuous %
and $\mathcal{G}(q) \sim \mathcal{G}_{pert}(q)=\frac{4\pi^2\gamma_m}{q^2\log (\frac{q^2}{\mu^2})}$ as $q^2\rightarrow \infty$.\footnote{QED also fulfills similar requirements if the Planck scale is taken as upper limit.} We define a simplest extension $\mathcal{G}_{simp}(q)=\frac{4\pi^2\gamma_m}{\max(q^2,\mu^2)\log (e+\frac{q^2}{\mu^2})}$ as a toy example. %

\section{Close to and below the critical point}
\subsection{Non-existence bound}
Following \cite{Infante2015}, we define
\begin{equation}
    f \equiv  \frac{\sup_{k,u}f(u,k)}{\sup_k u} = 1
\end{equation}
and with 
\begin{equation}
    K_{r} \equiv \sup_{p}\int_0^{\infty} dk K(k,p) %
\end{equation} 
we have
\begin{equation}
\lambda \equiv \frac{\sup_{p}T(u)}{\sup_p u} \leq  \frac{\sup_{k,u}f(u,k)}{\sup_p u}\sup_{p}\int_0^{\infty} dk K(k,p) = K_{r}
\end{equation}
and therefore, $K_{r}<1$ implies that there is no non-trivial solution for at least the one value of $p$ where $u$ is maximal. We get the most restrictive bound by finding $\inf_{r(p)} K_r$. 
For the above-mentioned simplest extension, we for example get $\gamma_m^{crit}\geq0.66/\max(Z^2(p))$ achieved at $p\sim 1.4\mu$ and $r(p) = (\mu^2+p^2)^{0.9}$ suitably adjusted to have the correct limit, such as $r(p) = (10+\log^{\delta}(\sqrt{e+\frac{p^2}{\mu^2}}))(\mu^2+p^2)^{0.9}/(10000\mu^2+p^2)^{0.9}$.

In taking the renormalization scale to infinity in the equation, we have lost the Wigner solutions as expected. %
Since we only consider the supremum, the same analysis applies to input functions which are allowed to be negative. Therefore, there are no Nambu solutions at all for $K_r<1$. 

\subsection{Bound on small solutions}
Next, assume $\inf_{a<p<b} u = \rho$ for some $a<b$. We define 
\begin{equation}
    K_r(a,b) \equiv \inf_{a<p<b}\int_a^{b} dk K(k,p)
\end{equation} 
so that we have
\begin{equation}
\begin{split}
\lambda(a,b) \equiv \inf_{a<p<b}\frac{T(u)}{\rho}
&\geq \inf_{a<p<b}\int_a^{b} dk\, K(k,p)\frac{f(u,k)}{\rho} \\
&\geq \inf_{a<k<b,\,u} \frac{f(u,k)}{\rho}
   \inf_{a<p<b}\int_a^{b} dk\, K(k,p)
   \geq \frac{K_r(a,b)}{1+\frac{\sup_{a<p<b} Z^2(p)\rho^2}{a^2 \inf_{a<p<b} r^2(p)}}
\end{split}
\end{equation}

$\sup_{a^2>\rho\mu,b,r(p)} K_r(a,b)>(1 + d\frac{\rho}{\mu}) \xrightarrow{\rho\rightarrow 0} 1$
for $d = \frac{\sup_{a<p<b} Z^2(p)}{ \inf_{a<p<b} r^2(p)}$ implies that there is no non-trivial infinitesimal solution for at least the one value of $a<p<b$ where $u$ is minimal. Therefore, we can place a bound on the appearance of any second-order DCSB phase transition.
For example, for the above-mentioned toy example, we get $\gamma_m^{crit}\leq1.0/\min(Z^2(p))$ with $a\sim0$, $b=4\mu$ and $r(p) = (\mu^2+p^2)^{0.9}$ again suitably adjusted.

Meanwhile, as long as $\sup_{a,b,r(p)}K_r(a,b)>1$, $T(u)(p)> u(p)$ for infinitesimal $\displaystyle\inf_{a<p<b} u(p)$, which includes all cases where $u$ is infinitesimal. %

\subsection{Combining the bounds \label{combining}}

The most optimal choice of $r(p)$ will be such that $\int_{a}^bdkK(k,p)$ is independent of $a<p<b$ because suppose $p^*$ is the location of the infimum. Then $r(p^*)$ can be raised without affecting the integral, showing that a better choice of $r(p)$ exists. 

We therefore want to show that given $p_{\infty}$, $r(p)$ can be chosen so that $\int_0^{\infty}dkK(k,p)$ is independent of $p$ for $p<p_{\infty}$ and decreasing thereafter. This is true iff $K_r = \int_{0}^\infty dkK(k,p)$ for $p<p_{\infty}$ has a valid solution for $r(p)$ and $K_r$,\footnote{One can always choose $p_{\infty}$ large enough that the integral is decreasing for $p>p_{\infty}$.} so define $t(p)=\frac{K_r}{r(p)}$. Expanding $K(k,p)$, we have
\begin{equation}
    t(p) = \frac{1}{K_r}\int_0^{\infty}dk \frac{k Z^2(k)}{4\pi^4}\int_{S^3}\mathcal{G}(q(k,p))t(k) \equiv \frac{1}{K_r}T_c(t)
\end{equation}

If we can define a total cone $K$ in a Banach space $E$ such that $T_c:K\rightarrow K$ is compact (we will do this in Sect.\ref{rigorous}), then by the Krein–Rutman theorem,\footnote{$T_c$ is ideal irreducible because it satisfies $T_c(u)(p)>0$ for all $p$ and all non-zero $u\in P$.} $T_c$ must have a largest eigenvalue $\lambda>0$ corresponding to a positive eigenvector $\bar t(p)$, therefore for $K_r=\lambda$, $\bar t(p)$ is a solution with asymptotic $\frac{1}{\log^{\gamma_m/K_r}\frac{p}{\mu}}$. For $K_r>1$, $\frac{\gamma_m}{K_r}<\gamma_m$, so that $r(p)$ satisfies the correct asymptotic as $p_{\infty}\rightarrow \infty$. Calling this solution $\bar r(p)$, we have $\sup_{a,b,r(p)}K_r(a,b) = K_{\bar r}$ and $\inf_{r(p)} K_r = K_{\bar r}$ by similar arguments. Then as $K_{\bar r}\rightarrow 1_+$, if a solution exists, we must have $\|u\|_{\infty}\rightarrow 0$ by dominated convergence and there is no solution for $K_{\bar r} < 1$.

\section{Proving existence past the critical point}

\subsection{Krasnosel'skii-Guo Theorem for $B(p)$ \label{rigorous}}

In order to utilize the Krasnosel'skii-Guo Theorem, we need to augment our space with a norm. Writing $\|u\|_{\infty} \equiv \sup_p u(p)$, consider therefore the Banach space $P_{\gamma_m}=\{B(x) \in C[(-\infty,\infty)]\ |\ B(x)\geq0, \lim_{x\rightarrow \infty}\frac{\partial \log B}{\partial \log x}=-\gamma_m\}$ with norm $\|u\| = \|u\|_{\infty} + \sup_{\lambda>1,p>\mu} \log^{\gamma_m-\delta} (\lambda) u(\lambda p)$. 
The purpose of the second piece will soon become apparent but it can be checked that this is indeed a Banach space. In particular, to check that the norm remains finite for $\lambda \rightarrow \infty$, we use the asymptotic bound %
$\lim_{k\rightarrow \infty}u(k) \sim \frac{m}{\log^{\gamma_m-\delta}\frac{k}{\mu}}$ so that $\lim_{\lambda\rightarrow \infty} \log^{\gamma_m-\delta} (\lambda) u(\lambda p) = m$. %
Let $P_{\Delta}$ be the $P_{\gamma_m}$ corresponding to $\gamma_m=0$ and take one choice of $B=B_0(p)$ with asymptotic mass $m$. %
Clearly, $P_{\Delta}$ is a total cone in $E$ and for any $B \in P_{\gamma_m}$, $B-B_0 \in P_{\Delta}$ with $\|B_0\|_{\infty}<\epsilon$ for some small $\epsilon$. We denote the corresponding $u_0(p)=B_0(p)r(p)$ and $T_{\Delta}(u) \equiv T(u+u_0)-u_0$.

$K(\cdot,p)$ is $L^1$-continuous by Scheff\'e's lemma which gives equicontinuity and then compactness of $T_{\Delta}$ under the $\|u\|_{\infty}$ part of the norm via the Arzelà–Ascoli theorem. For the second part of the norm, the same follows for each $\lambda$ after a variable transformation $s=p\lambda$.%

Let $P_{\Delta}^{r,R}$ be the annulus of $P_{\Delta}$ with $r<\|u\|<R$. %
To prove that $T_{\Delta}:P^{r,R}_{\Delta}\rightarrow P_{\Delta}$ past the critical point, %
we can use the same argument as for small $\|u\|_{\infty}$, $\|T(u)\|_{\infty}>\|u\|_{\infty}$, to show that $T(u)(p)>0$ for small $\|u\|_{\infty}$. For larger $\|u\|_{\infty}$, we have $T_{\Delta}(u) \geq \frac{1}{2}\left(\int_0^{\|u\|} dkK(k,p)\frac{k^2}{\|u\|_{\infty}}\right)+\int_{\|u\|}^{\infty}dkK(k,p)u_0(k)-u_0(p)$. If we require $\|u_0\|$ such that $2\|u_0\|_{\infty} R< \mu^2$ and $u_0(p)=0$ for $p<\mu$, then $\frac{k^2}{2\|u\|_{\infty}}>u_0(p)$ for $k<\|u\|_{\infty}$. Therefore, $T_{\Delta}(u)(p)\geq T_{\Delta}(0)(p)>0$. %

Consider $P_{\Delta,*}\subset P_{\Delta}$ the subspace of decreasing functions $u$ with $T_{\Delta}:P^{r,R}_{\Delta,*}\rightarrow P_{\Delta}$ and consider the boundary $\|u\|_{\infty}=R$ for $R\rightarrow \infty$. We have $\lim_{u\rightarrow \infty} f(u,k) \sim \frac{k^2}{u}$, therefore for given $k$, either $u(k)$ does not go to infinity with $\|u\|_{\infty}$, or it does and contributes as $\frac{k^2}{u}$ to the integral. %
This motivates that for $c=1$, $\lim_{\|u\|_{\infty}\rightarrow \infty}\frac{\|T_{\Delta}(u)\|_{\infty}}{\|u\|_{\infty}} = c \lim_{\|u\|_{\infty}\rightarrow \infty}\|\int_{\|u\|_{\infty}}^{\infty}dkK(k,p) \frac{u(k)}{\|u\|_{\infty}}\|_{\infty} = c \lim_{\|u\|_{\infty}\rightarrow \infty}\int_{\|u\|_{\infty}}^{\infty}dkK(k,\|u\|_{\infty}) \frac{u(k)}{\|u\|_{\infty}} %
$\footnote{$\mathcal G(q^2)\sim \mathcal G_{pert}(q^2)$ as $q^2\rightarrow \infty$ implies there exists a decay scale $q^*$ such that $\lim_{k^*\rightarrow \infty}\int_{k^*}^{\infty}dk (K(k^*,k)-K_{pert}(k^*,k))\sim \frac{(q^*)^2}{(k^*)^2}$ so for $k^*\gg q^*$, $K_{pert}(p,k)$ can be used. If $p< k^*$, the argument still applies as $q$ is increased by $k^*-p$ and if $p>k^*$, then the integral instead decays as $\frac{(q^*)^2}{p^2}$, which is even faster.} which is subtle because $\lim_{k^*\rightarrow \infty}\int_{k^*}^{\infty}dkK(k^*,k)=\frac{\gamma_m}{\delta}>1$ so as $\|u\|_{\infty}$ is scaled up, $u$ has to at the same time avoid stretching in momentum in order to avoid this limit. One can show that if $u(p)$ extends to no more than %
$p^*=\mu \left(\frac{\|u\|_{\infty}}{\mu}\right)^{\left(\frac{1}{1-n}\right)^{\frac{1}{\delta}}}$ and is zero afterwards, then less than $n\frac{\gamma_m}{\delta}$ is picked up with equality if $u(p)$ is effectively a Heaviside-function.\footnote{We have $c=1$ because for large $k,u$, corrections to $\frac{f(u)}{u}=1$ are suppressed by $\frac{\|u\|_{\infty}^2}{k^2}$ whereas $n\frac{\gamma_m}{\delta}$ is picked up over momenta depending on very large powers of $\frac{\|u\|_{\infty}}{\mu}$ for $\delta \sim 0$ and superlinear powers for all $\delta$, therefore $f(u,k)=u$ is a good approximation and for large $k<\|u\|_{\infty}$, corrections due to the region before $k=\|u\|_{\infty}$ are likewise suppressed by $\frac{k^2}{\|u\|_{\infty}^2}$. Therefore, $\|u\|_{\infty}$ as the minimum of integration is also valid.}

This is where the second part of the norm comes in. Suppose $u(p)$ barely decays up to $p^*$ and is zero afterwards, then %
$\|u\|=\|u\|_{\infty}(1+\log^{\gamma_m-\delta}\frac{p^*}{\mu}) = \|u\|_{\infty}(1+\left(\frac{1}{1-n}\right)^{\frac{\gamma_m-\delta}{\delta}}\log^{\gamma_m-\delta}(\frac{\|u\|_{\infty}}{\mu}))$. %
For $\|T_{\Delta}(u)\|$, we also have to include $\lim_{\|u\|\rightarrow \infty}\sup_{p>\mu,\lambda>1}\log^{\gamma_m-\delta}(\lambda)\int_{\|u\|_{\infty}}^{\infty}dk K(k,\lambda p) \sim \log^{\gamma_m-\delta}(\frac{\|u\|_{\infty}}{\mu})\frac{\gamma_m}{\delta}$ so that $\|T(u)\| \sim n \frac{\gamma_m}{\delta}(1+\log^{\gamma_m-\delta}(\frac{\|u\|_{\infty}}{\mu}))\|u\|_{\infty}$. We therefore have \begin{align}
    \sup_{0<n<1}\lim_{\|u\|\rightarrow \infty}\frac{\|T_{\Delta}(u)\|}{\|u\|} &= \sup_{0<n<1}\lim_{\|u\|\rightarrow \infty}n \frac{\gamma_m}{\delta}\frac{\|u\|_{\infty}}{\|u\|}(1+\log^{\gamma_m-\delta}(\frac{\|u\|_{\infty}}{\mu}))\\
    &= \sup_{0<n<1}\lim_{\|u\|\rightarrow \infty}\frac{n \frac{\gamma_m}{\delta}(1+\log^{\gamma_m-\delta}(\frac{\|u\|_{\infty}}{\mu})}{(1+\left(\frac{1}{1-n}\right)^{\frac{\gamma_m-\delta}{\delta}}\log^{\gamma_m-\delta}(\frac{\|u\|_{\infty}}{\mu}))}\\&=\sup_{0<n<1}n(1-n)^{\frac{\gamma_m-\delta}{\delta}}\frac{\gamma_m}{\delta}=\left(1-\frac{\delta}{\gamma_m}\right)^{\frac{\gamma_m}{\delta}-1}
\end{align}
and $\frac{1}{e}<\left(1-\frac{\delta}{\gamma_m}\right)^{\frac{\gamma_m}{\delta}-1}<1$ for $0 < \delta < \gamma_m$ where if $\|u\|_{\infty}$ stays finite, $n\rightarrow 1$ and the supremum is zero. For a more general $u(p)$, %
we first consider the sum of two barely decaying step-functions with $u=u_1+u_2$,  $\|u_2\|_{\infty}=\xi \|u_1\|_{\infty}$ and $n_2=\eta n_1$ for $\eta<1$. As $\|u\|\rightarrow \infty$, we have $(1+\xi)\|u_1\|\rightarrow \infty$ but if either $\|u_1\|\sim \|u\|$ or $\xi \|u_1\|\sim \|u\|$ does not hold, we are effectively back to a single step-function. We use $f(u)\leq f(u_1)+f(u_2)$ and the triangle inequality for $\|T_{\Delta}(u)\|$ and linearity of the norm for decreasing $u_{1,2}$ for $\|u\|$ so that 
\begin{align}
    &\sup_{\substack{0<n_1<1 \\ 0<\eta<1,\xi>0}}\lim_{\|u\|\rightarrow \infty}\frac{\|T_{\Delta}(u)\|}{\|u\|}\leq  \\&  \sup_{\substack{0<n_1<1 \\ 0<\eta<1,\xi>0}}\lim_{\|u\|\rightarrow \infty}\frac{n_1 \frac{\gamma_m}{\delta}(1+\log^{\gamma_m-\delta}(\frac{\|u_1\|_{\infty}}{\mu})+\eta \xi +\eta\xi \log^{\gamma_m-\delta}(\xi\frac{\|u_1\|_{\infty}}{\mu}))}{1+\xi + \left(\frac{1}{1-n_1}\right)^{\frac{\gamma_m-\delta}{\delta}}\log^{\gamma_m-\delta}(\frac{\|u_1\|_{\infty}}{\mu})+ \xi \left(\frac{1}{1-\eta n_1}\right)^{\frac{\gamma_m-\delta}{\delta}}\log^{\gamma_m-\delta}(\xi\frac{\|u_1\|_{\infty}}{\mu})}\\
    &=\sup_{\substack{0<n_1<1 \\ 0<\eta<1,\xi>0}}\lim_{\|u\|\rightarrow \infty}
    n_1(1-n_1)^{\frac{\gamma_m-\delta}{\delta}}\frac{\gamma_m}{\delta}\frac{\log^{\gamma_m-\delta}(\frac{\|u_1\|_{\infty}}{\mu})+\eta\xi \log^{\gamma_m-\delta}(\xi\frac{\|u_1\|_{\infty}}{\mu})}{\log^{\gamma_m-\delta}(\frac{\|u_1\|_{\infty}}{\mu})+\xi \left(\frac{1-n_1}{1-\eta n_1}\right)^\frac{\gamma_m-\delta}{\delta}\log^{\gamma_m-\delta}(\frac{\xi\|u_1\|_{\infty}}{\mu})}\\
    & =\sup_{\substack{0<n_1<1 \\ 0<\eta<1,\xi>0}}n_1(1-n_1)^{\frac{\gamma_m-\delta}{\delta}}\frac{\gamma_m}{\delta}\frac{1+\eta\xi}{1+\xi \left(\frac{1-n_1}{1-\eta n_1}\right)^\frac{\gamma_m-\delta}{\delta}} = \left(1-\frac{\delta}{\gamma_m}\right)^{\frac{\gamma_m}{\delta}-1}
\end{align}
because the function is strictly decreasing in $\eta$. 
For arbitrarily many step-functions, 
\begin{equation}
    \sup_{\substack{0<n_1<1 \\ 0<\eta_i<1,\xi_i>0}}\lim_{\|u\|\rightarrow \infty}\frac{\|T_{\Delta}(u)\|}{\|u\|}\leq\sup_{\substack{0<n_1<1 \\ 0<\eta_i<1,\xi_i>0}}n_1(1-n_1)^{\frac{\gamma_m-\delta}{\delta}}\frac{\gamma_m}{\delta}\frac{1+\sum_i\eta_i\xi_i}{1+\sum_i\xi_i \left(\frac{1-n_1}{1-\eta_i n_1}\right)^\frac{\gamma_m-\delta}{\delta}}
    \end{equation}
Since the top is linear in the $\eta_i$ and the bottom is convex increasing, the expression is Schur-concave and maximized by $\eta_i=0$ for all $i$ and yields the same result. 
Therefore, $\frac{1}{e}<\lim_{R\rightarrow \infty}\sup_{\|u\|=R}\frac{\|T_{\Delta}(u)\|}{\|u\|}< 1$ for general $u(p)$.

Since we already showed that past the critical point, $\|T_{\Delta}(u)\|>\|u\|$ for solutions with $\|u\|=r$ and $r\rightarrow 0$, we can now use the Krasnosel'skii-Guo Cone Compression Theorem: Define $\Omega^{a}$ the set of decreasing functions in $E$ with $\|u\|<a$. Then $P_{\Delta}\cap (\bar \Omega^{R}\setminus \Omega^r) = P^{r,R}_{\Delta,*}$ since there are no constant functions in $P_{\Delta}$ other than 0. Since $T_{\Delta}:P^{r,R}_{\Delta,*}\rightarrow P_{\Delta}$ is compact and satisfies the compressive boundary conditions, then $T_{\Delta}(u)$ has a fixed point in between the two boundaries, $r \leq \|u\|\leq R$. This proves that there exists a solution for each choice of $m\geq 0$, proves the DCSB transition is second-order by section \ref{combining} and moreover, that the solution $u(p)$ has to be decreasing. $\|T_{\Delta}(u)\|>\|u\|$ was proven for a particular choice of $r(p)$, and we need $\bar r(p)$ increasing to conclude that $B(p)$ is decreasing. Otherwise, there exists a different critical point over all increasing functions $r^*(p)$ above which solutions $B(p)$ are decreasing. For the model of \cite{Qin_2011} and $Z(p)=1$, $\bar r(p)$ is increasing and so the two critical points are the same.

Taking the gluon mass function and coupling from the fourth column of Table 1 in \cite{Qin_2011} for the rainbow-ladder vertex as a proxy for QCD, we find $T_c$'s largest eigenvalue equals one %
at $\frac{D}{\omega^2} \approx 0.7$ %
at $a=0$, $b=18$ GeV where we took $Z(p) = 1$ at the onset of DCSB breaking. In fact, the region from $12$ GeV to 18 GeV barely contributes. We find $\frac{D}{\omega^2}\approx 1$ in Figures in \cite{Wang_2012} with the discrepancy attributable to the assumption $Z(p)=1$. The simplest extension above breaks DCSB at $\gamma_m\approx 0.74$ for $Z_c(p) = 1$.

We now extend this result to the coupled equation for $M(p)$ and $Z(p)$ simultaneously.

\subsection{Schauder Theorem for $Z(p)$ and coupled equations}

The gap equation for $Z(p)$ is given by Eqn.\ref{TZ-eqn}. We require $K_Z(k,p)$ to be $L^1$-continuous. %
Then uniform boundedness follows by the reciprocal nature of the gap equation for $A(p)$ for an input space $\{A(p)\in C[[0,\infty)]\ |\ A(p)>A_-(p)>0\}$ and therefore compactness by the Arzelà–Ascoli theorem. %
If, as in the models in \cite{Qin_2011}, %
the kernel $K_Z(k,p)$ is negative for $k<k^*(p)$ and positive for $k>k^*(p)$ for each $p$ and we define $T_{\Delta}: P^{r,R}_{\Delta,*}\rightarrow P_{\Delta}$ as before with $u(p)\equiv M(p)r(p)$ and $r(p)$ increasing, then because $M(p)$ is decreasing, the finite mass term will suppress the negative part of the kernel $K^-_Z$ more than the positive part and push $T(A)$ from $T(A_0)$ in the direction of $A=1$ so that for $A(p)\geq \phi_0(p)\equiv\min[1,A_0(p)]$, we have $\phi_0(p)\leq T_Z(A)(p)\leq T^+_Z(\phi_0)(p)$ %
for finite $M(p)$ where $T^+_Z$ is the positive part of the kernel (likewise for $T^-_Z$ so that $T_Z = T_Z^++T_Z^-$). Therefore, we can use a combined Krasnosel'skii-Schauder theorem that has been derived for coupled systems\cite{INFANTE2024104165} as long as $\sup_{r} K_{r} > 1$ for $A=T_Z^+(\phi_0)$. In fact, for this value a solution for $A(p)$ to $M=0$ can also be guaranteed by noticing that if $A_-(p)$ can be found such that $A(p)>A_-(p)$ and $A_{--}(p)\equiv((T_Z^+)^2+T^-_Z) (A_-) >A_-(p)$, then $A_{--}(p)<T(A)(p)<T^+_Z(A_{-})(p)$,\footnote{Sufficient condition for $A_-(p)$ to be a lower bound and therefore $A_+=T^+_Z(A_-)$ to be an upper bound is $T^-_Z(A_-) + T_Z^+(A_+)>A_-$ for all $p$.} guaranteeing a solution for $M=0$ by Schauder's theorem and to the coupled system as long as $\sup_r K_r>1$ for $A(p)=T^+_Z(\min[1,A_{-}(p)])$. Therefore, we have $(T_{\Delta}, T_Z): P^{r,R}_{\Delta,*} \times P_Z \rightarrow  P_{\Delta} \times P_Z$ with $T_{\Delta}$ allowing us to use the Krasnosel'skii-Guo Theorem as before\footnote{The version used in \cite{INFANTE2024104165} uses a more general formulation due to Benjamin which however includes all cases where the Krasnosel'skii-Guo Theorem holds.} and $P_Z=\{A\in C[[0,\infty)]\ | \ A_-(p) \leq A(p) \leq T_Z^+(A_-)(p)\}$ a closed, convex and bounded set under sup-norm which is closed under $T_Z$, therefore amenable to Schauder's Theorem. Therefore, a positive, continuous solution pair $(u(p),A(p))$ with $u(p)$ decreasing must exist for every value of $m$ by the combined theorem. We require $\sup_{r^*} K_r > 1$ for $A(p)=T^+_Z(\min[1,A_{-}(p)])$ and $r^*(p)$ increasing so that $r(p)$ can be chosen to be increasing and $M(p)$ is decreasing.

For example, for the toy model, a solution is first guaranteed for $\gamma_m=1.1$. For the model in \cite{Qin_2011}, past the critical point at $\frac{D}{\omega^2} = 0.89$, such an $A_-(p)>0$ can be found at the physical point $\frac{D}{\omega^2} = 2.4$ and at points we checked for $1.7<\frac{D}{\omega^2}<3.6$. In each case, $\sup_{r^*} K_{r}>1$ so that $M(p)$ is decreasing. %
In \cite{Wang_2012}, a finite upper limit to $\frac{D}{\omega^2}$ was also found. The suspected reason is that increasing $\frac{D}{\omega^2}$ first raises but then lowers the eigenvalue for $A=T^+(\mathbf{1})$ whereas raising $\gamma_m$ always raises it.

\section*{Conclusion}
In conclusion, we studied Nambu solutions of the QCD RL gap equation at infinite renormalization scale. We found that for all positive, asymptotically perturbative kernels almost everywhere continuous in $L^1$%
, the critical point occurs when $T_c$'s largest eigenvalue $\lambda_{\max} = 1$ and satisfies $M(p)\rightarrow 0_+$ pointwise as $Z(p)\rightarrow Z_c(p)$ and $\mathcal G(q^2) \rightarrow \mathcal G_c(q^2)$. %
For all $\mathcal G$ for which $\lambda_{\max} < 1$, no non-trivial Nambu solution can exist and for $\lambda_{\max} > 1$, a positive, continuous and decreasing $u(p)$ must exist for every choice of $m\geq0$. If $\sup_{r^*(p),a,b} K_{r}(a,b)>1$ for $r^*(p)$ increasing, then $B(p)$ is also decreasing.

We extended our existence proof to the coupled system of equations for all $m\geq0$ if $ A_-(p)>0$ can be found such that for $M(p)=0$, $((T_Z^+)^2+T^-_Z)  A_-(p) >  A_-(p)$ and $\lambda_{\max}>1$ for $A=T_Z^+(\min[1,A_{-}(p)])$ with $\sup_{r^*(p),a,b} K_{r}(a,b)>1$ so that $M(p)$ is decreasing. %
For the model of the fourth column of Table 1 of \cite{Qin_2011}, this is the case for the physical point $\frac{D}{\omega^2}=2.4$ and for individual points we checked for $1.75<\frac{D}{\omega^2}<3.6$, where $\frac{D}{\omega^2}=0.89$ is the critical point.

Considering a more general vertex, for example the Ball-Chui vertex\cite{Roberts_1994,PhysRevD.22.2542}, we encounter serious difficulties with this approach due to the fact that a finite-difference derivative term $A'(p)$ appears in the equation for $B(p)$ which is not constrained to have any sign or magnitude by Schauder's Theorem. Concerning proving $M(p)\rightarrow 0^+$ for a more general vertex, the Krein-Rutman theorem relies on a positive kernel which is no longer guaranteed.%

Part of the issue with quantum field theory is that the infinite tower of Dyson-Schwinger equations may not have a guaranteed solution space. A corollary to the results in this paper is that for Nambu solutions to the gap equation in QCD, the contractive condition at $\|u\|\rightarrow \infty$ for the Krasnosel'skii-Guo theorem is satisfied for any $m$. %

\subsection*{Acknowledgements}
We thank C.D. Roberts %
for useful feedback.

\bibliography{mybib}

\appendix
\section*{Appendix}
\section{Gap equation primer for mathematicians \label{A}}

The gap equation is a 2-component coupled integral equation for the quark mass function $M(p)$ and field renormalization function $Z(p)$ as a function of 4-momentum $p_{\mu}$ with $p \equiv \sqrt{-p^2}$ in terms of the gluon propagator times coupling $g^2(q)D_{\mu \nu}(q)$ and vertex $\Gamma_{\mu}(p,k,q)$ with $k_{\mu}$ and $q_{\mu}=p_{\mu}-k_{\mu}$ loop momenta:

\begin{equation}
S^{-1}(p)\equiv\frac{1}{Z(p)}(\slashed p - M(p)) = \slashed p - m(\mu_{ren}) -\Sigma(p)+\Sigma(\mu_{ren})
\end{equation}

with \begin{equation}
\Sigma(p) = iZ_1\frac{4}{3}\int \frac{d^4 q}{(2\pi)^4} \gamma_{\mu} g^2(q)D^{\mu\nu}(q)S(k)\Gamma_{\nu}(k,p)
\end{equation}

The simplification of the rainbow-ladder vertex amounts to $\Gamma_{\mu}(p,k)=\gamma_{\mu}$ and we use Landau gauge so that $Z_1g^2(q)D_{\mu\nu}(q) = (g_{\mu\nu}-\frac{q_{\mu}{q_{\nu}}}{q^2})\mathcal G(q)$\cite{Qin_2011}. After rotating to the Euclidean metric and letting $ A(p)\equiv \frac{1}{Z(p)}$ and $B(p)\equiv \frac{M(p)}{Z(p)}$, we have the operators \cite{Roberts_1994,Aguilar_2007} 
\begin{equation}
    T(A,B)  = \frac{1}{4\pi^4}\int_0^{\infty} dk k^3\frac{B(k)}{A^2(k)k^2+B^2(k)} \int_{S^3}\mathcal{G}(q^2(k,p))
    \label{T-eqn}
\end{equation}
and 
\begin{equation}
    T_Z(A,B)  = 1+\frac{1}{12\pi^4 p^2}\int_0^{\infty}dk k^3 \frac{A(k)}{A^2(k)k^2+B^2(k)} \int_{S^3} \mathcal G(q^2(k,p)) \left(p\cdot k + 2\frac{p\cdot q k\cdot q}{q^2}\right)
    \label{TZ-eqn}
\end{equation}
with $S^3$ the surface of the sphere that leaves $k^2$ and $p^2$ invariant. Then the trace of the gap equation reads $B(p) = m(\mu_{ren}) + T(A,B)(p)-T(A,B)(\mu_{ren})$ and the trace of $\slashed p$ times the gap equation reads $ A(p) = 1+T_Z(A,B)(p)-T_Z(A,B)(\mu_{ren})$. Here $\mu_{ren}$ is the renormalization scale which is an experimentally related quantity: Since we cannot perform experiments at infinite momentum, we cannot determine a function by its exact asymptotic behavior. Instead, the experimental scale $\mu_{ren}$ is used where a measurement of the mass $m(\mu_{ren})$ has been made. However, since we know perturbative, that is large momentum functions are accurate, we can let $\mu_{ren}\rightarrow \infty$, in which case we simply have $B(p) = T(A,B)(p)$ and $ A(p) = T_Z(A, B)(p)$. Solutions are degenerate and will be identified by a continuous parameter $m$ which is equivalent to some $m(\mu_{ren})$ at an arbitrary scale $\mu_{ren}$.\footnote{In the QED and Higgs sector, interactions don't flow to zero at high energies so that a high-energy cutoff replaces $\mu_{ren}$ as we send $\mu_{ren}\rightarrow \infty$. In fact, viewed from this high-energy viewpoint, the standard model parameters are extremely fine-tuned for life\cite{uzan2025}.}%

By letting $\mu_{ren}\rightarrow \infty$, we lose solutions which require $m(\mu_{ren})$ to achieve an eigenvalue of $1$, so-called Wigner solutions. %
The remaining solutions are called Nambu solutions and some of the time come in pairs for the RL vertex, as $M\rightarrow -M$ is a symmetry of the gap equation broken only by $m$. The positive Nambu solutions are considered physical minima given positive asymptotic masses, the negative Nambu solutions saddle points with negative pion-mass squared and the Wigner solutions maxima with negative pion-mass squared as well as sigma-mass squared\cite{Wang_2012}.

\section{Fixed-point theorem primer for physicists \label{B}}
The version of the Krasnosel'skii theorem we use is the compressive part of the formulation adapted by Guo in which we start with a Banach space $E$ and cone $K\subset E$. A Banach space is a complete, normed vector space with norm $\|u\|$, a cone consists of a set of functions $u$ as well as their scaled copies $\lambda u$ for any $\lambda \geq 0$ such that no function except zero has its additive inverse in the space. We require that the cone is closed and convex: convexity requires any weighted average of two functions to be within the space and closure includes the boundary. Defining bounded, open sets $\Omega$ and $V$, let $T(u)$ be a compact map from $K\cap (\bar \Omega \setminus V)$ into $K$. Compact maps are identified by the Arzelà–Ascoli theorem if they map bounded sets to sets that are bounded and equicontinuous - $L^1$-continuous in our case. %
Then if we have $\|T(u)\|<\|u\|$ for $u\in \partial \Omega \cap K$ and $\|T(u)\|>\|u\|$ for $u\in \partial V \cap K$,
then $T(u)=u$ must have a solution for some $u\in K\cap (\bar \Omega \setminus V)$.%

The Schauder Fixed Point Theorem states that if $P_Z$ is a convex, closed and bounded subset of $E$ and $T_Z$ a bounded, compact map, then if $T_Z:P_Z\rightarrow P_Z$, $T_Z(u)=u$ has a solution inside $P_Z$.

A total cone is a convex cone $K$ such that $K \cup -K$ is dense in $E$. Then the Krein-Rutman Theorem implies that a compact operator $T:K\rightarrow K$ which is ideal irreducible has an eigenvector in $K$ whose eigenvalue is $T$'s largest (the so-called spectral radius).

\section{Tail of the quark mass function}

Let's extract the large-momentum behavior of the quark mass function. In this limit, the dominant contribution comes from $\mathcal G(q) \sim \frac{4\pi^2\gamma_m}{q^2 \log (\frac{q^2}{\mu^2})}$.

We can borrow a result of the angular integral
\begin{equation}
    \int_{S^3} \frac{1}{q^2}= 4\pi \min\left[\frac{\pi}{2p^2},\frac{\pi}{2k^2}\right]
\end{equation}
to estimate
\begin{equation}
    \int_{S^3} \frac{1}{q^2 \log (\frac{p^2}{\mu^2})}= 4\pi\min\left[\frac{\pi}{2p^2 \log (\frac{p^2}{\mu^2})},\frac{\pi}{2k^2 \log (\frac{k^2}{\mu^2})}\right] + \chi(k,p)
\end{equation}
For the tail of the mass function, the integral over $\chi(k,p)$ can be ignored both in and away from the chiral limit, %
as we will show after solving for both. Then removing the renormalization scale by taking it to infinity, we obtain for $p\gg\mu$
\begin{equation}
    \frac{M(p)}{Z(p)} = 2\gamma_m\int^{\infty}_0 dk k^3 \frac{Z(k)M(k)}{k^2+M(k)^2} \min\left[\frac{1}{p^2 \log \frac{p^2}{\mu^2}},\frac{1}{k^2 \log \frac{k^2}{\mu^2}}\right] %
\end{equation}
and with $\frac{M(p)}{Z(p)} = B(p)$%
, we get
\begin{equation}
    \left(B'(p)%
p^3 \log \frac{p^2}{\mu^2}(1+\frac{1}{\log \frac{p^2}{\mu^2}})\right)' = -4\gamma_m p Z^2(p)B(p) 
\label{differential}
\end{equation}

To leading order in $\log \frac{p^2}{\mu^2}$, we have $Z(p)=1$ 
and using $x = \log \frac{p}{\mu}$, we get
\begin{equation}
    \lim_{x\rightarrow \infty}B(x) = \frac{c_1}{x^{\gamma_m}} + e^{-2x}c_2 x^{\gamma_m-1} \equiv c_1 B_+(x)+c_2B_-(x)
\end{equation}

For $B_+(p)$, to leading order in $\frac{1}{x}$ it is sufficient to have the region $k>p$ such that
\begin{equation}
    B(p) = 2 \gamma_m \int_p^{\infty} dk k B(k)\frac{1}{k^2\log \frac{k^2}{\mu^2}}
\end{equation}
leading to 
\begin{equation}
    B'(p) = -2 \gamma_m p B(p)\frac{1}{p^2\log \frac{p^2}{\mu^2}}
\end{equation}
which is solved by
\begin{equation}
    B(x) = \frac{m}{x^{\gamma_m}}
\end{equation}

For $B_-(p)$, the integral domain $[p,\infty]$'s contribution is likewise suppressed by $\frac{1}{x}$. Therefore, define a scale $\Lambda$ such that the asymptotic form is valid above $\Lambda$ up to $\frac{1}{x}$ corrections. Call $\Delta_{\Lambda} \equiv 2\gamma_m\int^{\Lambda}_0 dk k^3 \frac{Z(k)M(k)}{k^2+M(k)^2}$. Then we have for $p>\Lambda$
\begin{equation}
    B_-(x)=  \frac{\Delta_{\Lambda}}{p^2\log\frac{p^2}{\mu^2}} + \frac{2\gamma_m}{p^2\log\frac{p^2}{\mu^2}}\int^{x}_{x_{\Lambda}} dy B_-(y)e^{2y}
\end{equation}
with $x_{\Lambda} = \log\frac{\Lambda}{\mu}$ %
and we arrive at 
\begin{equation}
    c_2 = (2x)^{-\gamma_m}(\Delta_{\Lambda}-c_2(2x_{\Lambda})^{\gamma_m}) +  c_2 
\end{equation}
The solution is 
\begin{equation}
    c_2 = \Delta_{\Lambda}(2x_{\Lambda})^{-\gamma_m}
\end{equation}

Finally, we integrate $B_{\pm}(p)$ over $\chi(k,p)$ and find that it is %
$0.02/\log^2(\frac{p^2}{\mu^2})$ suppressed over the other term for $p>4\mu$.

\section{Fixed point analysis}
\subsection{Away from the chiral limit \label{C}}
For any input function $B(x)$ and its limit function $B_+(p)$ %
in a series expansion around $x=\infty$, let $j^*$ be the first non-zero term of $B(p)/B_+(p)$: $B(x) = B_+(x)(1+\frac{a_{j^*}}{x^{j^*}} + \ldots)$. %
The integral domain $[0,p]$'s contribution is suppressed by $\frac{1}{x}$ over the region $[p,\infty]$. Therefore, $T(B)$ can be expanded as $T(B)(x) = B_+(x)(1+\frac{a'_{j^*}}{x^{j^*}} + \ldots)$ %
with $a'_{j^*} = \frac{\gamma_m}{\gamma_m+j^*}a_{j^*}$. Since $\frac{\gamma_m}{\gamma_m+j^*}<1$ for any positive $j^*$, $B_+$ is an attractive fixed-point which proves that current quark mass cannot be dynamically generated so that if $(\gamma_m,m)$ characterizes $B(p)$, then it also characterizes $T(B)(p)$.%

\subsection{In the chiral limit}
With the same definitions as in the section above and with the same %
$j^*$ while ignoring the integral domain $[p,\infty]$ which is suppressed by $\frac{1}{x}$, if $j^*<2\gamma_m$, then $T(B)$ can be expanded as $T(B)(x) = B_-(x)(1+\frac{a'_{j^{*}}}{x^{j^{*}}} + \ldots)$ with $a'_{j^{*}} = \frac{\gamma_m}{\gamma_m-j^*}a_{j^*}$. Since $|\frac{\gamma_m}{\gamma_m-j^*}|>1$ for $j^*<2\gamma_m$, $B_-$ would at least be repulsive in one iteration. Meanwhile, for $j^*>2\gamma_m$, $B_-$ is an attractive fixed point.%

\section{Tail of the quark field renormalization function}

Following Eqn.\ref{TZ-eqn} with $\mathcal G(q) \sim \frac{4\pi^2\gamma_m}{q^2 \log (\frac{q^2}{\mu^2})}$ as the dominant contribution, we observe that
\begin{equation}
    \int_{S^3} \frac{p\cdot k+2 p\cdot q k\cdot q/q^2}{q^2}= 0
\end{equation}
but we still find
\begin{equation}
    \int_{S^3} \frac{p\cdot k+2 p\cdot q k\cdot q/q^2}{q^2\log\frac{q^2}{\mu^ 2}}= 3\pi^2\min\left[\frac{k^2}{p^2 \log^2 \frac{p^2}{\mu^2}},\frac{p^2}{k^2 \log^2 \frac{k^2}{\mu^2}}\right] +\chi_Z(k,p)
\end{equation}
with $\chi_Z(k,p)$ polynomially suppressed over the other term at large $k\gg p$. Therefore, for $A(p)$ converging to $1$ in the UV as required by the renormalization condition at $\infty$, the integral converges due to the square of the log in the denominator and the renormalization scale can be taken to infinity in the gap equation. However, after converting to differential form as in Eqn.\ref{differential}, because of the $x^2$ factor, $\frac{1}{x^n}$ perturbations grow to $\frac{1}{x^{n-1}}$ in one iteration, making the impact of $\chi_Z(k,p)$ unpredictable. We prove in the main text that there exist $A_{1,2}$ such that if $A_1(p)<A(p)<A_2(p)$, then the same is true for $T(A)$ which circumvents this issue.

\end{document}